\documentclass[final]{ws-ijmpd}

\newcommand{\D}{\hat{D}}
\begin{document}

\markboth{A. B. Balakin, H. Dehnen and A. E. Zayats} {Nonminimal
isotropic cosmological model with Yang-Mills and Higgs fields}

\catchline{}{}{}{}{}
%
%%%%%%%%%%%%%%%%%%%%%%%%%%%%%%%%%%%%%%%%%%%%%%%%%%%%%%%%%%%%%%%%%%%%

\title{NONMINIMAL ISOTROPIC COSMOLOGICAL MODEL WITH YANG-MILLS AND
HIGGS FIELDS}

\author{ALEXANDER B. BALAKIN}

\address{Department of General Relativity and Gravitation,\\ Kazan State University,\\ Kremlevskaya str. 18,
Kazan, 420008, Russia\\
Alexander.Balakin@ksu.ru}

\author{HEINZ DEHNEN}

\address{Fachbereich Physik, \\ Universit\"at Konstanz,\\ Fach M 677, D-78457, Konstanz, Germany\\
Heinz.Dehnen@uni-konstanz.de}

\author{ALEXEI E. ZAYATS}

\address{Department of General Relativity and Gravitation,\\ Kazan State University,\\ Kremlevskaya str. 18,
Kazan, 420008, Russia\\
Alexei.Zayats@ksu.ru}

\maketitle

\begin{abstract}
We establish a nonminimal Einstein-Yang-Mills-Higgs model, which
contains six coupling parameters. First three parameters relate to
the nonminimal coupling of non-Abelian gauge field and gravity
field, two parameters describe the so-called derivative nonminimal
coupling of scalar multiplet with gravity field, and the sixth parameter
introduces the standard coupling of scalar field with Ricci scalar.
The formulated six-parameter nonminimal
Einstein-Yang-Mills-Higgs model is applied to cosmology. We show
that there exists a unique exact cosmological solution of the de
Sitter type for a special choice of the coupling parameters. The
nonminimally extended Yang-Mills and Higgs equations are satisfied
for arbitrary gauge and scalar fields, when the coupling parameters
are specifically related to the curvature constant of the isotropic
spacetime. Basing on this special exact solution we discuss the
problem of a hidden anisotropy of the Yang-Mills field, and give an
explicit example, when the nonminimal coupling effectively screens
the anisotropy induced by the Yang-Mills field and thus restores the
isotropy of the model.
\end{abstract}


\section{Introduction}\label{Intro}

The discussion of a nonminimal coupling (NMC) of gravity
with fields and media has a long history. The most intensely this
topic has been studied  in connection with the problem of nonminimal
coupling of gravity and scalar field, which has numerous cosmological
applications. The details of the investigations of this problem are
discussed, e.g., in the review of
Faraoni {\it et al\/}.\cite{FaraR} The development of the theory of NMC
of gravity and scalar field $\phi$ has started by the introduction
of the term $\xi \phi^2 R$ to the Lagrangian ($R$ is the Ricci
scalar). In Ref.~\refcite{Chernikov} the special choice $\xi = 1/6$ has been
motivated by the conformal invariance; in Ref.~\refcite{Callan} this
quantity was considered as an arbitrary parameter of the model. Such
a model has been widely used for the cosmological applications, in
which $\xi$ played a role of extra parameter of inflation (see,
e.g., Refs.~\refcite{Abbott}--\refcite{Fara4}). In
Refs.~\refcite{HDehnen1}--\refcite{HDehnen4} the gauge-invariant
term $\alpha {\bf \Phi}^{+}{\bf \Phi}R $ has been introduced instead
of $\xi \phi^2 R$ in the context of non-Abelian gauge theory (${\bf
\Phi}$~is a multiplet of scalar complex Higgs fields interacting
with gravity and spinor matter.)
Subsequent generalizations have been related to the replacement of $\xi
\phi^2$ by the function $f({\Phi}^2)$ (see, e.g.,
Refs.~\refcite{Bergmann}--\refcite{Steinh}), as well as, to the
inserting of the terms of the type $F({\Phi}^2, {\cal R})$ both linear
and nonlinear in the Ricci scalar, Ricci and Riemann tensors
(see, e.g., Refs.~\refcite{Linde}--\refcite{Inagaki}). The idea of
nonminimal derivative coupling introduced in Ref.~\refcite{Amen3} and
developed further in Refs.~\refcite{Capo1,Capo2} has enriched
the NMC modeling by the terms $\phi_{,ij..}$.
Nonminimal cosmological models based on the formalism of derivative coupling
are the multi-parameter ones and have supplementary
abilities for a fitting of observational data. Let us note that the
NMC of gravity and scalar field leads to the modifications of both the
Klein-Gordon and the Einstein equations, and such modifications are
of interest for various inflation scenarios.
Thus, the modeling of nonminimal interactions of scalar and
gravitational fields is one of the well established and physically
motivated branch of modern cosmology. Natural extension of the
nonminimal theory from the models with scalar fields coupled to
curvature to the models describing scalar fields interacting with
gauge fields has the same sound motivation and can disclose new
aspects of cosmological dynamics.

The study of the nonminimal coupling of gravity with electromagnetic
field has another motivation and another history. In 1971
Prasanna\cite{Prasa1} introduced the invariant $R^{ikmn}F_{ik}F_{mn}$
($R^{ikmn}$ is the Riemann tensor, $F_{ik}$ is the Maxwell tensor)
as a possible element of a Lagrangian, and then in
Ref.~\refcite{Prasa2} obtained the corresponding nonminimal
one-parameter modification of the Einstein-Maxwell equations. In
1979 Novello and Salim\cite{Novello1}  proposed to insert the gauge
non-invariant terms $R A^k A_k$ and $R^{ik}A_i A_k$ in the
Lagrangian ($A_k$ is an electromagnetic potential four-vector). A
qualitatively new step has been made by Drummond and
Hathrell in Ref.~\refcite{Drum}, where the one-loop corrections to the quantum
electrodynamics (QED) are obtained, which take into account the nonminimal coupling
of gravity and electromagnetism. The Lagrangian of
such a theory happens to contain three fundamental
$U(1)$-gauge-invariant scalars $R^{ikmn}F_{ik}F_{mn}$,
$R^{ik}g^{mn}F_{im}F_{kn}$ and $RF_{mn}F^{mn}$ with  coefficients
reciprocal to the square of the electron mass. This Lagrangian had
no arbitrary parameters, but curvature induced modifications of the electrodynamic
equations gave the impetus to wide discussions about the
formal structure of the nonminimal Lagrangian, basic evolutionary equations,
breaking the conformal invariance and
the properties of the photons, coupled to curvature in  different
gravitational backgrounds (see, e.g.,
Refs.~\refcite{Acci1}--\refcite{Lafrance}).
The last paper revived, as well, the interest to the paradigm:
curvature coupling and equivalence principle, various aspects of which are
now discussed (see, e.g., Refs.~\refcite{Prasa3,Solanki}).
The QED-motivation of the use of
the generalized Maxwell equations can also be found in the papers of
Kosteleck\'y and colleagues.\cite{Kost1,Kost2} The effect of
birefringence induced by curvature, first discussed in
Ref.~\refcite{Drum}, and some of its consequences for the
electrodynamic systems have been investigated
in Refs.~\refcite{Balakin1}--\refcite{Balakin5}
for the case of pp-wave background. The
generalization of the idea of nonminimal interactions to the case of
torsion coupled to the electromagnetic field has been made in
Refs.~\refcite{Hehl1,Hehl2} (see, also, Ref.~\refcite{Hehl3} for a
review on the problem).
To summarize we stress that the study of electrodynamic systems
nonminimally coupled to the gravity field poses a natural question about
curvature induced variations of photon velocity in the cosmological
background. Since the interpretation of observational data in
cosmology depends essentially on the velocity of photon propagation
during different cosmological epochs, the modeling of nonminimal
electrodynamic phenomena seems to be well motivated and interesting
from physical point of view.

Concerning the nonminimal Einstein-Yang-Mills (EYM) theory, we can
distinguish between two different ways to establish it. The first way
is the direct nonminimal generalization of the Einstein-Yang-Mills (EYM) theory.
In the framework of this approach Horndeski\cite{Horn} and
M\"uller-Hoissen\cite{MH} obtained the nonminimal one-parameter EYM model from a
dimensional reduction of the Gauss-Bonnet action. Now the
Gauss-Bonnet models are of great interest in connection with the
problem of dark energy (see, e.g., the Gauss-Bonnet model with
nonminimal scalar field\cite{OdinDE}).
Thus, the non-Abelian multi-parameter
extensions of nonminimal models are also well motivated, since they give a chance
to explain the accelerated expansion of the Universe without addressing to exotic substance.
We follow the alternative way, which is connected with a non-Abelian
generalization of the nonminimal Einstein-Maxwell theory  along the
lines proposed by Drummond and Hathrell\cite{Drum} for the linear
electrodynamics. Based on the results of Ref.~\refcite{BL05}
a three-parameter gauge-invariant nonminimal EYM model
linear in curvature is considered.\cite{1BZ06}\cdash\cite{BDZ07}
Our goal is to formulate a nonminimal Einstein-Yang-Mills-Higgs
(EYMH) theory, and this process, of course, also admits different
approaches. In fact, the nonminimal EYMH theory should accumulate
the ideas and methods both from the nonminimally extended EYM theory
and from the nonminimally extended scalar field theory. Initial
attempt to develop nonminimal EYMH theory can be found, for
instance, in Ref.~\refcite{Bij}, where the scalar Higgs field is
nonminimally coupled with gravity via $\xi {\Phi}^2 R$ term, and the
Higgs field ${\bf \Phi}$ is included into the Lagrangian of the
Yang-Mills field in a composition with a square of the Yang-Mills
potential: ${\Phi}^2 A_k^{(a)} A^k_{(a)}$. Such a theory is not
gauge-invariant.

In this paper we establish a new six-parameter nonminimal
Einstein-Yang-Mills-Higgs model. First three coupling
parameters, $q_1$, $q_2$ and $q_3$, describe a nonminimal interaction
of Yang-Mills field and gravitational field. The fourth and fifth
parameters, $q_4$ and $q_5$, describe the so-called gauge-invariant
nonminimal derivative coupling of the Higgs field with gravity.
Since the gauge-invariant derivative, $\D_m {\Phi}^{(a)}$, contains
the potential of the Yang-Mills field, the corresponding nonminimal
term is associated with ``triple'' interaction, namely,
gravitational and scalar fields, gauge and scalar fields, and gauge
and gravitational fields. The sixth parameter, $\xi$, is the
well-known coupling parameter nonminimally connecting gravitational
and scalar fields via the term $\xi R {\Phi}^2$. Of course, this
model is only one of a wide class of the nonminimal
EYMH models.
As for its motivation and possible physical
applications, one can see that on the one hand, the interest to a
six-parameter nonminimal EYMH model is based on the sound results
obtained earlier in the framework of partial nonminimal models
(Einstein-Maxwell, Einstein-Yang-Mills and scalar field theories),
on the other hand, the six-parameter model under discussion shows
new specific solutions of cosmological type, which can not appear in
more simple models.

The paper is organized as follows. In Sec. \ref{Formalism} we
formulate the nonminimal EYMH model, which contains six
phenomenological coupling parameters, and establish the nonminimally
extended Yang-Mills, Higgs and Einstein equations. In Sec.
\ref{IsModel} we apply the introduced master equations to the
spacetime with constant curvature and obtain the specific
relationships between coupling constants, which turn the extended
equations for the gauge field and scalar field into identities. In
Subsec. \ref{dSsptime} we discuss the exact solutions to the
nonminimal EYMH equations attributed to the isotropic cosmological
model with Yang-Mills field, characterized by hidden anisotropy.

\section{The formalism of the nonminimal EYMH
theory}\label{Formalism}

\subsection{Minimal EYMH theory and basic definitions}\label{MinEYMH}

The minimal Einstein-Yang-Mills-Higgs theory can be
formulated in terms of the action functional
\begin{equation}
S_{(\rm EYMH)} = \int d^4 x \sqrt{-g} \left\{\frac{R}{\kappa} +
\frac12 F_{mn}^{(a)} F^{mn}_{(a)} - \D_m \Phi^{(a)} \D^m \Phi_{(a)}
+ {V}({\Phi}^2) \right\} , \label{act}
\end{equation}
where $g = {\rm det}(g_{ik})$ is the determinant of a metric tensor
$g_{ik}$, $R$ is the Ricci scalar, Latin indices run from 0 to 3.
The symbol ${\Phi}^{(a)}$ denotes the multiplet of the Higgs scalar
fields, ${V}({\Phi}^2)$ is a potential of the Higgs field and
$\Phi^2\equiv\Phi^{(a)}\Phi_{(a)}$. Let us mention that there are
two formal variants to introduce the cosmological constant into the
action (\ref{act}): first, explicitly as an additional term
$\frac{2\Lambda}{\kappa}$, second, as a term $V(0)$ in the
decomposition
\begin{equation}
{V}({\Phi}^2)= \frac{2\Lambda}{\kappa} + \mu {\Phi}^2 + \omega
{\Phi}^4 + \dots \label{V}
\end{equation}
Below we consider the second variant. Following
Ref.~\refcite{Rubakov}, Section 4.3, we consider the Yang-Mills
field ${\bf F}_{mn}$ and the Higgs field ${\bf \Phi}$ taking values
in the Lie algebra of the gauge group $SU(n)$:
\begin{equation}
{\bf F}_{mn} = - i {\cal G} {\bf t}_{(a)} F^{(a)}_{mn} \,, \quad
{\bf A}_m = - i {\cal G} {\bf t}_{(a)} A^{(a)}_m \,, \quad {\bf
\Phi} = {\bf t}_{(a)} \Phi^{(a)} \,. \label{represent}
\end{equation}
Here ${\bf t}_{(a)}$ are the Hermitian traceless generators of
$SU(n)$ group, the constant ${\cal G}$ is the strength of the gauge
coupling, $F^{(a)}_{mn}$, $A^{(a)}_m$ and $\Phi^{(a)}$ are real
fields ($A^{(a)}_m$ represents the Yang-Mills field potential) and
the group index $(a)$ runs from $1$ to $n^2-1$.
The symmetric tensor $G_{(a)(b)}\equiv 2\, {\rm Tr} \ {\bf t}_{(a)}
{\bf t}_{(b)}$ plays a role of a metric in the group space so that,
e.g., $\Phi_{(a)}\equiv G_{(a)(b)}\Phi^{(b)}$.
The Yang-Mills fields $F^{(a)}_{mn}$ are connected with the
potentials of the gauge field $A^{(a)}_i$ by the well-known formula
(see, e.g., Refs.~\refcite{Rubakov}--\refcite{Akhiezer})
\begin{eqnarray}
F^{(a)}_{mn} = \nabla_m A^{(a)}_n - \nabla_n A^{(a)}_m + {\cal G}
f^{(a)}_{\ (b)(c)} A^{(b)}_m A^{(c)}_n \,. \label{Fmn}
\end{eqnarray}
Here $\nabla _m$ is a  covariant spacetime derivative, the symbols
$f^{(a)}_{\ (b)(c)}$ denote the real structure constants of the
gauge group $SU(n)$. The gauge-invariant derivative is defined
according to the formula (Ref.~\refcite{Rubakov}, Eqs.(4.46, 4.47))
\begin{equation}
\D_m \Phi^{(a)} \equiv \nabla_m \Phi^{(a)} + {\cal G} f^{(a)}_{\
(b)(c)} A^{(b)}_m \Phi^{(c)} \,. \label{DPhi}
\end{equation}
For the derivative of arbitrary tensor defined in the group space
we use the following rule\cite{Akhiezer}:
\begin{eqnarray}
\D_m Q^{(a) \cdot \cdot \cdot}_{\cdot \cdot \cdot (d)} \equiv
\nabla_m Q^{(a) \cdot \cdot \cdot}_{\cdot \cdot \cdot (d)} + {\cal
G} f^{(a)}_{\ (b)(c)} A^{(b)}_m Q^{(c) \cdot \cdot \cdot}_{\cdot
\cdot \cdot (d)} - {\cal G} f^{(c)}_{\ (b)(d)} A^{(b)}_m Q^{(a)
\cdot \cdot \cdot}_{\cdot \cdot \cdot (c)} + \dots \label{DQ2}
\end{eqnarray}
The commutator and anticommutator of the generators ${\bf t}_{(a)}$
take the form
\begin{equation}
\left[ {\bf t}_{(a)} , {\bf t}_{(b)} \right] = i  f^{(c)}_{\ (a)(b)}
{\bf t}_{(c)} \,, \label{fabc}
\end{equation}
\begin{equation}
\left\{ {\bf t}_{(a)} , {\bf t}_{(b)} \right\} \equiv {\bf t}_{(a)}
{\bf t}_{(b)} + {\bf t}_{(b)} {\bf t}_{(a)} = \frac{1}{n}\,
G_{(a)(b)} {\bf I} + d^{\,(c)}_{\ (a)(b)} {\bf t}_{(c)}\,,
\label{dabc}
\end{equation}
where $d_{(c)(a)(b)}$ are the completely symmetric coefficients and
${\bf I}$ is the unitary matrix.
The metric $G_{(a)(b)}$, the structure constants $f^{(c)}_{\ (a)(b)}$
and the coefficients $d^{\,(c)}_{\ (a)(b)}$ are supposed to be
constant tensors in standard and covariant
manner\cite{Akhiezer}. This means that
\begin{gather}
\partial_m G_{(a)(b)} = 0 \,,\quad
\partial_m  f^{(a)}_{\ (b)(c)} = 0 \,, \quad \partial_m d^{\,(a)}_{\ (b)(c)} =
0\,, \nonumber\\ \D_m G_{(a)(b)} = 0 \,,\quad \D_m f^{(a)}_{\
(b)(c)} = 0 \,, \quad \D_m d^{\,(a)}_{\ (b)(c)} = 0\,. \label{DfG}
\end{gather}

\subsection{Nonminimal extension of the Lagrangian}\label{Nmextension}

Any version of nonminimal generalization of the Lagrangian of the EYMH
theory is based on the choice of the set of admissible invariants.
The classification of the
Yang-Mills fields based on the invariant polynomials in the
$F^{(a)}_{ik}$ tensor has been made in Ref.~\refcite{Roskies}.
Nonlinear constitutive equations for Yang-Mills field along a line
of the Born-Infeld theory has been first discussed in
Ref.~\refcite{Mills} (see, also, e.g., Ref.~\refcite{Goldin}).
Possessing the tensorial quantities $F^{(a)}_{ik}$, $\Phi^{(a)}$,
$\D_i \Phi^{(a)}$, $R^{ikmn}$, $R^{ik}$, $R$ one can construct a
variety of gauge invariant scalars both minimal and nonminimal. This
procedure has been discussed in the framework of scalar field theory
and electrodynamics (see Introduction and references therein).
Einstein-Yang-Mills-Higgs theory possesses an extended set of basic
elements for such a representation, thus, a number of candidates to
be included into a Lagrangian is much bigger. For instance, in order
to couple the group indices $(a)$ and $(b)$ in the product
$F^{(a)}_{ik} F^{(b)}_{mn}$, we can use, first, the standard
convolution procedure, based on metric $G_{(a)(b)}$, second, the
projections onto $\Phi_{(a)} \Phi_{(b)}$, $\Phi_{(a)} \D_j
\Phi_{(b)}$ or $\D_j \Phi_{(a)} \D_s \Phi_{(b)}$,  third, the
convolution with symmetric tensors $d_{(a)(b)(c)} \Phi^{(c)}$,
$d_{(a)(b)(c)} \D_j \Phi^{(c)}$, or with antisymmetric tensors
$f^{(c)}_{ \ (a)(b)} \Phi_{(c)}$, $f^{(c)}_{ \ (a)(b)} \D_j
\Phi_{(c)}$. The corresponding examples of the scalar invariants,
admissible for including into the nonminimal Lagrangian are
$$
\frac{1}{2} {\cal R}^{ikmn}_{(I)} F^{(a)}_{ik} F^{(b)}_{mn} \left[
G_{(a)(b)} + d_{(a)(b)(c)} \Phi^{(c)} \Psi_1( {\Phi}^2) +
\Phi_{(a)} \Phi_{(b)} \Psi_2( {\Phi}^2)\right.
$$
\begin{equation}
\left. {}+(\D_l \Phi_{(a)}) (\D^l \Phi_{(b)}) \Psi_3( {\Phi}^2)+
d_{(a)(b)(c)} \Phi^{(c)} (\D^l \Phi_{(h)}) (\D_l \Phi^{(h)})
\Psi_4( {\Phi}^2) + \dots\right] \,, \label{inv1}
\end{equation}
\begin{equation}
{\cal R}^{ikmn}_{(II)} F^{(a)}_{ik} \left[ f_{(a)(b)(c)}(\D_m
\Phi^{(b)}) (\D_n \Phi^{(c)}) \Psi_5( {\Phi}^2) + \dots\right] \,,
\label{in1}
\end{equation}
$$
{\cal R}^{ikmn}_{(III)} \left\{ g_{im} (\D_k \Phi^{(a)}) (\D_n
\Phi^{(b)}) \left[ G_{(a)(b)} {+} d_{(a)(b)(c)} \Phi^{(c)} \Psi_6(
{\Phi}^2) {+} \Phi_{(a)} \Phi_{(b)} \Psi_7( {\Phi}^2) {+}
\dots\right] \right.
$$
\begin{equation}
\left.  {}+(\D_i \Phi^{(a)})(\D_k \Phi^{(b)})(\D_m
\Phi^{(c)})(\D_n \Phi^{(d)}) f^{(h)}_{\cdot (a)(b)} f_{(h)(c)(d)}
\Psi_8( {\Phi}^2) + \dots \right\} \,. \label{in2}
\end{equation}
Here $\Psi_1,\ \Psi_2,\ \dots,\ \Psi_8$ are arbitrary functions of
their argument, and the tensors ${\cal R}^{ikmn}_{(I)}$, ${\cal
R}^{ikmn}_{(II)}$ and ${\cal R}^{ikmn}_{(III)}$ are considered to
be appropriate linear combinations of the Riemann tensor and
its convolutions with phenomenological coupling constants $q_1,\
q_2,\ \dots,\ q_j$. These constants are treated to be independent
and have a dimensionality of area. Below the examples of such
tensors are presented explicitly.

In this paper we restrict ourselves to the consideration of a Lagrangian,
which satisfy the following requirements: the EYMH Lagrangian is a
gauge invariant scalar linear in a spacetime curvature, quadratic
in the Yang-Mills field strength tensor $F^{(a)}_{ik}$ and depending
on the first derivative of the Higgs field only. In addition, in
this paper we consider the convolutions of the standard type only,
i.e., the terms including $G_{(a)(b)} {\mathstrut F}^{(a)}_{ik}
{\mathstrut F}^{(b)}_{mn}$, $G_{(a)(b)}\Phi^{(a)}\Phi^{(b)}$, etc.
We intend to consider more sophisticated models in future papers.

\subsection{Explicit example of nonminimal gauge-invariant
Lagrangian}\label{NMEYMH}

Consider now an action functional
\begin{eqnarray}\label{1act}
S_{({\rm NMEYMH})} = \int d^4 x \sqrt{-g}\ \left\{
\frac{R}{\kappa}+\frac{1}{2}F^{(a)}_{ik} F^{ik}_{(a)}
-{\D}_m\Phi^{(a)}{\D}^m\Phi_{(a)}+V(\Phi^2) \right.
{}\nonumber\\
\left. {}+\frac{1}{2} {\cal R}^{ikmn}F^{(a)}_{ik} F_{mn(a)} -
{\Re}^{\,mn}{\D}_m\Phi^{(a)}{\D}_n\Phi_{(a)}+\xi R\,{\Phi}^2
\right\}\,,
\end{eqnarray}
where the tensors ${\cal R}^{ikmn}$ and $\Re^{\,mn}$ are defined
as follows:
\begin{eqnarray}
{\cal R}^{ikmn} &\equiv&
\frac{q_1}{2}R\,(g^{im}g^{kn}-g^{in}g^{km}) \nonumber\\
{}&+& \frac{q_2}{2}(R^{im}g^{kn} - R^{in}g^{km} + R^{kn}g^{im}
-R^{km}g^{in}) + q_3 R^{ikmn}\,, \label{sus}
\end{eqnarray}
\begin{equation}\label{Re}
\Re^{\,mn}\equiv {q_4}Rg^{mn}+q_5 R^{mn}\,.
\end{equation}
This action describes a six-parameter nonminimal
Einstein-Yang-Mills-Higgs model, and $q_1$, $q_2,\ \dots,\ q_5$,
$\xi$ are the constants of nonminimal coupling.

\subsubsection{Nonminimal extension of the Yang-Mills field
equations}\label{Ymequations}

The variation of the action $S_{({\rm NMEYMH})}$ with respect to
the Yang-Mills potential $A^{(a)}_i$ yields
\begin{equation}
\D_k {H}^{ik}_{(a)}  =  - {\cal G} (\D_k \Phi^{(b)})f_{(a)(b)(c)}
\Phi^{(c)} \left( g^{ik} + \Re^{ik} \right) \,. \label{Heqs}
\end{equation}
Here the tensor ${H}^{ik}_{(a)}$  is defined as
\begin{equation}
{H}^{ik}_{(a)} = \left[ \frac{1}{2}( g^{im} g^{kn} - g^{in} g^{km})
+ {\cal R}^{ikmn} \right] G_{(a)(b)} F^{(b)}_{mn} \,. \label{HikR}
\end{equation}
This equation looks like Maxwell equation for the medium with the
susceptibility tensor ${\cal R}^{ikmn}$ and the current vector
${\cal G} (\D_k \Phi^{(b)})f_{(a)(b)(c)} \Phi^{(c)} \left( g^{ik} +
\Re^{ik} \right)$ induced by the Higgs field.

\subsubsection{Nonminimal extension of the Higgs field equations}\label{Hequations}

The variation of the action $S_{({\rm NMEYMH})}$ with respect to
the Higgs scalar field $\Phi^{(a)}$ yields
\begin{equation}\label{Heq}
{\D}_m\left({\D}^m{\Phi}^{(a)} + \Re^{\,mn}\D_n{\Phi^{(a)}}\right) =
- \xi{R\,{\Phi}^{(a)}}-V'(\Phi^2){\Phi^{(a)}} \,.
\end{equation}
This equation can be rewritten in the form
\begin{equation}\label{21Heq}
{\D}_m {\Psi}^{m(a)} = - \left[ \xi R + V'(\Phi^2) \right]
{\Phi}^{(a)} \,, \quad {\Psi}^{m(a)} \equiv {\D}^m{\Phi}^{(a)} +
\Re^{\,mn}\D_n{\Phi}^{(a)} \,,
\end{equation}
and can be considered as scalar analog of (\ref{Heqs}) and
(\ref{HikR}).

\subsubsection{Master equations for the gravitational field}\label{Einequations}

In the nonminimal theory linear in curvature the equations for the
gravity field related to the action functional $S_{({\rm NMEYMH})}$
take the form
\begin{equation}
\left(R_{ik}-\frac{1}{2}Rg_{ik}\right)\cdot(1+\kappa\xi\Phi^2)=
\kappa\xi\left(\D_i\D_k-g_{ik}\D_m\D^m\right){\Phi}^2+
{k}T^{(NMYMH)}_{ik} \,. \label{Eeq}
\end{equation}
The principal novelty of these equations, in comparison with the
well-known equations for nonminimal scalar field, is associated with
the third, fourth, etc., terms in the decomposition
\begin{equation}
T^{(NMYMH)}_{ik}=T^{(YM)}_{ik} + T^{(H)}_{ik} + q_1 T^{(I)}_{ik} +
q_2 T^{(II)}_{ik} + q_3 T^{(III)}_{ik} + q_4 T^{(IV)}_{ik} + q_5
T^{(V)}_{ik} \,. \label{Tdecomp}
\end{equation}
The first term $T^{(YM)}_{ik}$:
\begin{equation}
T^{(YM)}_{ik} \equiv \frac{1}{4} g_{ik} F^{(a)}_{mn}F^{mn}_{(a)} -
F^{(a)}_{in}F_{k\,(a)}^{\ n} \,, \label{TYM}
\end{equation}
is a stress-energy tensor of pure Yang-Mills field. The second
one, $T^{(H)}_{ik}$,
\begin{equation}
T^{(H)}_{ik}=\D_i\Phi^{(a)}\D_k\Phi_{(a)}-\frac{1}{2}g_{ik}\D_m\Phi^{(a)}\D^m\Phi_{(a)}+\frac{1}{2}V(\Phi^2)\,g_{ik}
\end{equation}
is a stress-energy tensor of the Higgs field. The definitions of
other five tensors are related to the corresponding coupling
constants $q_1$, $q_2,\,\dots,\,q_5$:
\begin{equation}%
T^{(I)}_{ik} = R\,T^{(YM)}_{ik} -  \frac{1}{2} R_{ik}
F^{(a)}_{mn}F^{mn}_{(a)} + \frac{1}{2} \left[ {\D}_{i} {\D}_{k} -
g_{ik} {\D}^l {\D}_l \right] \left[F^{(a)}_{mn}F^{mn}_{(a)}
\right] \,, \label{TI}
\end{equation}%
$$%
T^{(II)}_{ik} = -\frac{1}{2}g_{ik}\biggl[{\D}_{m}
{\D}_{l}\left(F^{mn(a)}F^{l}_{\ n(a)}\right)-R_{lm}F^{mn (a)}
F^{l}_{\ n(a)} \biggr] $$%
$${}- F^{ln(a)} \left(R_{il}F_{kn(a)} +
R_{kl}F_{in(a)}\right)-R^{mn}F^{(a)}_{im} F_{kn(a)} - \frac{1}{2}
{\D}^m{\D}_m \left(F^{(a)}_{in} F_{k\,(a)}^{ \
n}\right)$$%
\begin{equation}%
\quad{}+\frac{1}{2}{\D}_l \left[ {\D}_i \left(
F^{(a)}_{kn}F^{ln}_{(a)} \right) + {\D}_k
\left(F^{(a)}_{in}F^{ln}_{(a)} \right) \right] \,, \label{TII}
\end{equation}%
$$
T^{(III)}_{ik} = \frac{1}{4}g_{ik} R^{mnls}F^{(a)}_{mn}F_{ls(a)}-
\frac{3}{4} F^{ls(a)} \left(F_{i\,(a)}^{\ n} R_{knls} +
F_{k\,(a)}^{\ n}R_{inls}\right)$$%
\begin{equation}%
\quad {}-\frac{1}{2}{\D}_{m} {\D}_{n} \left[ F_{i}^{ \ n
(a)}F_{k\,(a)}^{ \ m} + F_{k}^{ \ n(a)} F_{i\,(a)}^{ \ m} \right]
\,, \label{TIII}
\end{equation}%

$$%
T^{(IV)}_{ik}=\left(R_{ik}-\frac{1}{2}Rg_{ik}\right)\D_m\Phi^{(a)}\D^m\Phi_{(a)}+R\,\D_i\Phi^{(a)}\D_k\Phi_{(a)}
$$%
\begin{equation}\label{TIV}
    {}+\left(g_{ik}\D_n\D^n-\D_i\D_k\right)\left[\D_m\Phi^{(a)}\D^m\Phi_{(a)}\right]\,,
\end{equation}

$$%
T^{(V)}_{ik}=\D_m\Phi^{(a)}\left[R_i^m\D_k\Phi_{(a)}+R_k^m\D_i\Phi_{(a)}\right]-
\frac{1}{2}R_{ik}\,\D_m\Phi^{(a)}\D^m\Phi_{(a)}
$$%
$$%
{}+\frac{1}{2}g_{ik}\D_m\D_n\left[\D^m\Phi^{(a)}\D^n\Phi_{(a)}\right]-
\frac{1}{2}\,\D^m\biggl\{\D_i\left[\D_m\Phi^{(a)}\D_k\Phi_{(a)}\right]
$$%
\begin{equation}\label{TV}
    {}+\D_k\left[\D_m\Phi^{(a)}\D_i\Phi_{(a)}\right]-
    \D_m\left[\D_i\Phi^{(a)}\D_k\Phi_{(a)}\right]\biggr\}\,.
\end{equation}

\subsubsection{Bianchi identities}\label{Bids}

The Einstein tensor $R_{ik}-\frac{1}{2}g_{ik}R$ is the
divergence-free one, thus, the tensor $T^{(NMYMH)}_{ik}$ in the
right-hand-side of (\ref{Eeq}) has to satisfy the differential
condition
\begin{equation}
\nabla^k
\left\{\frac{\kappa\xi\left(\D_i\D_k-g_{ik}\D_m\D^m\right){\Phi}^2+
{k}T^{(NMYMH)}_{ik}}{(1+\kappa\xi\Phi^2)} \right\} =0 \,.
\label{Eeeq}
\end{equation}
One can prove that it is valid automatically, when $F^{(a)}_{ik}$ is
a solution of the Yang-Mills equations (\ref{Heqs}), and
$\Phi^{(a)}$ satisfy the Higgs equations (\ref{Heq}). In order to
check this fact directly, one has to use the Bianchi identities and
the properties of the Riemann tensor:
\begin{equation}
\nabla_i R_{klmn} + \nabla_l R_{ikmn} + \nabla_k R_{limn} = 0 \,,
\quad R_{klmn} + R_{mkln} + R_{lmkn} = 0 \,, \label{bianchi}
\end{equation}
as well as the rules for the commutation of covariant derivatives
\begin{equation}
(\nabla_l \nabla_k - \nabla_k \nabla_l) {\cal A}^i = {\cal A}^m
R^i_{\cdot mlk} \,, \label{nana}
\end{equation}
(this rule is written here for the vector only). The procedure of
checking is analogous to one, described in Ref.~\refcite{Acci3}
and we omit it.

\section{Isotropic cosmological model associated with six-parameter nonminimal EYMH theory}\label{IsModel}

Generally, the application of the EYMH model  to cosmological
problems requires the spacetime to be considered as anisotropic one.
Clearly, when the spacetime is isotropic, the Einstein tensor in the
left-hand-side of (\ref{Eeq}) is diagonal, while the tensor
$T_{ik}^{({\rm NMYMH})}$ in the right-hand-side is generally
non-diagonal. This can be also motivated by the analogy with
Einstein-Maxwell theory: it is well-known, for instance, that the
minimal models with magnetic field are inevitably anisotropic and
can be properly described in terms of Bianchi models. Nevertheless,
as it was shown in Ref.~\refcite{BZ05}, the nonminimal extension of
the Einstein-Maxwell theory admits the models in which the spacetime
is isotropic while the magnetic field is non-vanishing. Below we
discuss the first example of analogous problem in the framework of
nonminimal EYMH theory. Our goal is to present explicitly an exact
solution to the equations of spatially isotropic EYMH model. When
the Yang-Mills field is non-vanishing, the stress-energy tensor
(\ref{Tdecomp}) is non-diagonal, as in the case of Einstein-Maxwell
theory, thus, the symmetry of equations for the gravitational field
is, generally, broken. Nevertheless, we will indicate a special
choice of the coupling parameters $q_1$, $q_2,\ \dots,\ q_5$, $\xi$,
for which one can guarantee, that these equations become
self-consistent. Since the de Sitter model is associated with the
spacetime of constant curvature, we consider a number of properties
of desired solution without solving the master equations.

\subsection{Constant curvature spacetime \\ and restrictions on the Yang-Mills-Higgs fields}\label{ConstCurv}

We consider isotropic cosmological models with constant curvature $K$.\cite{MTW}
For these spacetimes the Riemann tensor takes the form
\begin{equation}\label{curv}
R_{ikmn}=-K\left(g_{im}g_{kn}-g_{in}g_{km}\right)
\end{equation}
and the Ricci tensor, the Ricci scalar are
\begin{equation}\label{1curv}
R_{ik}=-3Kg_{ik}\,, \qquad R=-12K\,.
\end{equation}
The tensors ${\cal R}_{ikmn}$ and $\Re_{ik}$, introduced
phenomenologically, can be transformed into
\begin{equation}\label{simpa}
{\cal
R}_{ikmn}=-K(6q_1+3q_2+q_3)\left(g_{im}g_{kn}-g_{in}g_{km}\right)\,,\quad
\Re_{ik}=-3K(4q_4+q_5)g_{ik}\,.
\end{equation}
Then, the ${H}^{ik}_{(a)}$ tensor and the ${\Psi}^m_{(a)}$ vector
simplify significantly, and the equations (\ref{HikR}) and
(\ref{21Heq}) convert, respectively, into
\begin{equation}\label{1simpa}
 [1-2K(6q_1+3q_2+q_3)]  \D_k {F}^{ik}_{(a)} =  -{\cal G}[1-3K(4q_4+q_5)] f_{(a)(b)(c)}\D^i \Phi^{(b)}\Phi^{(c)} \,,
\end{equation}
\begin{equation}\label{2simpa}
[1-3K(4q_4+q_5)] \D_m \D^m {\Phi}^{(a)} =  \left[ 12 \xi K -
V'(\Phi^2) \right] {\Phi}^{(a)} \,.
\end{equation}
We focus on the case, when the equation for the Yang-Mills field
turns into identity for arbitrary (non-vanishing) ${F}^{(a)}_{ik}$.
It is not possible, when the EYMH theory is minimal one.
Nevertheless, in the framework of nonminimal EYMH theory with
non-vanishing Higgs field, ${\Phi}^{(a)}\neq 0$, the Yang-Mills
equations admit an arbitrary non-vanishing solution, when
\begin{equation}\label{AQU}
2(6q_1+3q_2+q_3)= \frac{1}{K} \,, \quad 3(4q_4+q_5) = \frac{1}{K}
\,.
\end{equation}
If (\ref{AQU}) is valid, the Higgs equations are self-consistent,
when
\begin{equation}\label{condV}
\left[ 12 \xi K - V'(\Phi^2) \right] {\Phi^{(a)}} =0 \,.
\end{equation}
In its turn, it is possible in two cases: first, when
$V({\Phi}^2)$ is a linear function of its argument,
\begin{equation}\label{1condV}
V({\Phi}^2) = \frac{2\Lambda}{\kappa} + 12 K \xi {\Phi}^2 \,,
\end{equation}
${\Phi}^{(a)}$ being arbitrary, second, when ${\Phi}^2$ is constant
satisfying the equation (\ref{condV}).

\subsection{One-parameter nonminimal EYMH model}\label{1parModel}

In order to obtain an analytical progress in the searching for the
solution to the gravity field equations let us consider the
one-parameter model, which is characterized by the following
conditions:
\begin{equation}\label{q}
q_1=q_4=\frac{1}{12K} \,, \quad q_2=q_3=q_5=0 \,, \quad
V({\Phi}^2) = \frac{2\Lambda}{\kappa} + \mu {\Phi}^2 \,, \quad \xi
= \frac{\mu}{12K} \,.
\end{equation}
These conditions guarantee that the Yang-Mills equations
(\ref{Heqs}) and the Higgs equations (\ref{Heq}) are the trivial
identities for arbitrary ${F}^{ik}_{(a)}$ and ${\Phi}^{(a)}$. The
Einstein equations for this case take the form
$$%
3Kg_{ik} \ (1+\kappa\xi {\Phi}^2)=
\kappa\xi\left(\D_i\D_k-g_{ik}\D_p\D^p\right){\Phi}^2+
\frac{\kappa}{2}\,V({\Phi}^2)g_{ik}
$$%
$$%
{}+\frac{\kappa}{8}\,g_{ik}\left(F_{mn}^{(a)}F^{mn}_{(a)}-2\D_m\Phi^{(a)}\D^m\Phi_{(a)}\right)
$$%
\begin{equation}\label{EinSpec}
   {}+\frac{\kappa}{24K}\left(\D_i\D_k-g_{ik}\D_p\D^p\right)\left[F_{mn}^{(a)}F^{mn}_{(a)}-2\D_m\Phi^{(a)}\D^m\Phi_{(a)}\right]\,.
\end{equation}
It can be reduced formally to ten equations for one scalar
function
\begin{equation}\label{W}
\nabla_i \nabla_k W = K g_{ik} W \,,
\end{equation}
where
\begin{equation}\label{W1}
W \equiv
F_{mn}^{(a)}F^{mn}_{(a)}-2\D_m\Phi^{(a)}\D^m\Phi_{(a)}+24K\xi\Phi^2
+ \frac{24}{\kappa} \left(\frac{\Lambda}{3} - K \right) \,.
\end{equation}
Let us consider the integrability conditions for such system and
calculate the commutator of the covariant derivatives $\hat{{\cal
K}}_{ijk} \equiv [\nabla_i \nabla_j - \nabla_j \nabla_i] \nabla_k
W $. On the one hand with (\ref{W}) this commutator yields
directly
\begin{equation}\label{KK1}
\hat{{\cal K}}_{ijk} = - K\left(g_{ik}\nabla_j W - g_{jk}\nabla_i
W \right) \,.
\end{equation}
On the other hand due to (\ref{nana})
\begin{equation}\label{KK2}
\hat{{\cal K}}_{ijk} = -{R^p}_{kij}\nabla_p W = - K
\left(g_{ik}\nabla_j W - g_{jk}\nabla_i W\right) \,.
\end{equation}
Thus, the integrability conditions are satisfied identically, and
the equations (\ref{W}) are completely integrable.

\subsection{De Sitter spacetime}\label{dSsptime}

In order to represent the exact solution to (\ref{W})
explicitly we consider the model with positive curvature, $K>0$,
and reduce the metric to the de Sitter form\cite{Weinberg}
\begin{equation}\label{deSitter}
ds^2 = dt^2 + \exp\{2 \sqrt{K} t\} (\eta_{\alpha \beta}
dx^{\alpha} dx^{\beta}) \,,
\end{equation}
where $\alpha, \beta = 1,2,3$ and $\eta_{\alpha \beta}$ is the
spatial part of the Minkowski metric with the signature $(-,-,-)$.
Then (\ref{W}) splits into three subsystems
$$
\partial^2_t W -K W =0 \,, \quad \partial_{\alpha} [\partial_t W -\sqrt{K}
W]=0 \,,
$$
\begin{equation}\label{EqE}
\partial_{\alpha} \partial_{\beta} W + \eta_{\alpha \beta} \sqrt{K} \exp\{2 \sqrt{K} t\}
[\partial_t W -\sqrt{K} W] = 0 \,,
\end{equation}
which can be readily solved
\begin{equation}\label{SolEqE}
W =  C_1 e^{\sqrt{K}t} + C_2 e^{- \sqrt{K}t} + e^{\sqrt{K}t}
\left[L_{\alpha} x^{\alpha} + C_2 K \eta_{\alpha \beta} x^{\alpha}
x^{\beta} \right] \,.
\end{equation}
Here $C_1$, $C_2$ and $L_{\alpha}$ are arbitrary constants. Thus,
we obtain an exact solution of the total EYMH system of equations
for which the Yang-Mills field $F_{mn}^{(a)}$ and the Higgs fields
$\Phi^{(a)}$ are connected by unique condition
$$
F_{mn}^{(a)}F^{mn}_{(a)}-2\D_m\Phi^{(a)}\D^m\Phi_{(a)}+24K\xi\Phi^2
+ \frac{24}{\kappa} \left(\frac{\Lambda}{3} - K \right)
$$
\begin{equation}\label{W11}
\qquad{}= C_1 e^{\sqrt{K}t} + C_2 e^{- \sqrt{K}t} + e^{\sqrt{K}t}
\left[L_{\alpha} x^{\alpha} + C_2 K \eta_{\alpha \beta} x^{\alpha}
x^{\beta} \right] \,.
\end{equation}
Clearly, there exists a lot of various Yang-Mills-Higgs
configurations, which satisfy this condition.

\section{Discussions}\label{Discussion}
\noindent
\par 1. The main mathematical result of the presented paper is the
establishing of a new self-consistent nonminimal system of master
equations for the coupled Yang-Mills, Higgs and  gravity fields from
the gauge-invariant nonminimal Lagrangian (\ref{1act}). The obtained
mathematical model contains six arbitrary parameters, and, thus,
admits a wide choice of special sub-models interesting for the
applications to the nonminimal cosmology (isotropic and anisotropic)
and nonminimal colored spherical symmetric objects. The applications
require the phenomenological coupling constants $q_1$, $q_2,\
\dots,\ q_5$ and $\xi$ to be interpreted adequately. Following the
idea, discussed in Ref.~\refcite{HDehnen4}, we intend not to
introduce ``new constants of Nature'', but to relate the
phenomenological parameters with the constants well-known in the
High Energy Particle Physics, on the one hand, and with the
constants of cosmological origin, on the other hand. Indeed, in the
specific cosmological model, established above, the sixth
phenomenological parameter $\xi$ is expressed in terms of the square
of the effective mass of the Higgs bosons $\mu$ and constant
curvature $K$, $\xi = \frac{\mu}{12K}$. Other parameters are
expressed in terms of $K$ (see (\ref{q})). Since in the de Sitter
model the Hubble constant is $H=\sqrt{K}$, one can say that $q_1$,
$q_2,\ \dots,\ q_5$ are connected with $H$. Analogously, one can
consider the equality $H^2=K=\frac{\Lambda}{3}$ and thus, one can
say that they are connected with the cosmological constant
$\Lambda$. In any case the parameters of nonminimal coupling $q_1$,
$q_2,\ \dots,\ q_5$ can be expressed in terms of cosmological
parameters $K$, $H$ or $\Lambda$, and define a specific radius of
curvature coupling, $r_q \equiv \frac{1}{\sqrt{K}}$ and the corresponding
time parameter $t_q \equiv r_q/c$.

2. The curvature coupling modifies the master equations for the
Yang-Mills and Higgs fields. According to (\ref{Heqs}) a new tensor
${H}^{ik}_{(a)}$ appears (see (\ref{HikR})), which is an analog of
the induction tensor in the Maxwell theory\cite{Maugin}. This means
that the curvature coupling of the non-Abelian gauge field with
gravity acts as a sort of quasi-medium with a nonminimal
susceptibility tensor ${\cal R}^{ikmn}$ (see (\ref{HikR})). As well,
the curvature coupling modifies the master equations for the Higgs
field, and the tensor $\Re^{mn}$, according to (\ref{Heq}), can be
indicated as a simplest nonminimal susceptibility tensor for the
Higgs field, and the vector ${\Psi}^{m}_{(a)}$ (see (\ref{21Heq}))
can be defined as scalar induction. For the specific set of
coupling constants (see (\ref{AQU}), (\ref{q})) the non-Abelian induction
$H^{ik}_{(a)}$ and the scalar induction ${\Psi}^{m}_{(a)}$ can turn
into zero, despite the fact that the Yang-Mills field strength
${F}^{ik}_{(a)}$ and the Higgs field ${\Phi}^{(a)}$ are
non-vanishing. This means that, when (\ref{AQU}) holds, the
possibility exists to satisfy the nonminimally extended Yang-Mills
and Higgs equations for arbitrary ${F}^{ik}_{(a)}$ and
${\Phi}^{(a)}$.
This possibility gives, in principle, a new option for modeling
physical processes in Early Universe and shows very interesting
analogy between this nonminimal model and resonance phenomena in
plasma physics. Indeed, when we deal with plasma waves (for
instance, with the longitudinal waves) one can see that electric
induction $\vec{D}$ is connected with the longitudinal electric
field $\vec{E}_{||}$ with the frequency $\omega$ by the relation
$\vec{D}= \varepsilon_{||}\vec{E}_{||}$. Here $\varepsilon_{||}$ is
the longitudinal dielectric permittivity, the simplest expression
for this quantity can be obtained in the limit of long waves and
gives $\varepsilon_{||}= 1-\frac{\Omega^2_{p}}{\omega^2}$, where
$\Omega_{p}$ is the well-known plasma frequency. When
$\omega=\Omega_{p}$, one obtains $\vec{D}=0$ and electrodynamic
equations are satisfied for arbitrary $\vec{E}_{||}$. Analogous
feature can be found in the nonminimal model described above (see
Eq. (\ref{1simpa})). Indeed, the quantity $K$ with the dimensionality of
squared frequency ($c=1$) can be regarded as an analog of $\Omega^2_{p}$, the
quantity $2(6q_1+3q_2+q_3)$ can be indicated as $1/\omega^2$, then
the term $1-2K(6q_1+3q_2+q_3)$ plays a role of effective
permittivity scalar $\varepsilon_q$. When this effective permittivity scalar
vanishes, i.e., when the constants of nonminimal coupling are
connected with the constant curvature $K$ according to (\ref{AQU}), we obtain the
resonance case, for which the Yang-Mills and Higgs equations are
satisfied identically for arbitrary strength field tensor
$F^{ik}_{(a)}$ and Higgs multiplet $\Phi^{(a)}$, the color induction
$H^{ik}_{(a)}$ being equal to zero.

3. The vector potential of the Yang-Mills field $ A^{(a)}_i$ enters
the master equations via the gauge covariant derivative $\hat{D}_k$,
thus, the gauge field generates an anisotropy in the
spacetime. Such an anisotropy, in general case, breaks down the symmetry
of the model and produces the isotropy violation. Nevertheless, as
it was shown above, the nonminimal coupling can effectively screen
the anisotropy and guarantee the symmetry conservation. In the
framework of this model one can speak about hidden anisotropy of
the Yang-Mills field, keeping in mind that non-Abelian gauge field
enters the master equations for the gravity field in the isotropic
combinations only.

\section*{Acknowledgments}

This work was funded by DFG through the project No. 436RUS113/487/0-5.


\begin{thebibliography}{00}

\bibitem{FaraR} V. Faraoni, E. Gunzig and P. Nardone, {\it Fund.
Cosmic Phys.} {\bf 20}, 121 (1999).

\bibitem{Chernikov} N. A. Chernikov and E. A. Tagirov, {\it Ann. Inst. Henri
Poincar\'e} {\bf A9}, 109 (1968).

\bibitem{Callan} C. G. Callan, Jr., S. Coleman and R. Jackiw, {\it Ann. Phys. (N.Y.) } {\bf 59}, 42 (1970).

\bibitem{Abbott} L. F. Abbott, {\it Nucl. Phys.} {\bf B185}, 233 (1981).

\bibitem{Starob} A. A. Starobinski, {\it Sov. Astron. Lett.} {\bf
7}, 36 (1981).

\bibitem{Turn1} F. S. Accetta, D. J. Zoller and M. S. Turner, {\it Phys. Rev.} {\bf
D31}, 3046 (1985).

\bibitem{Futa} T. Futamase and K. Maeda, {\it Phys. Rev.} {\bf
D39}, 399 (1989).

\bibitem{Amen1} L. Amendola, M. Litterio and F. Occhionero, {\it Int. J. Mod. Phys.} {\bf A5}, 3861 (1990).

\bibitem{Fara4} V. Faraoni, {\it Phys. Rev.} {\bf D53}, 6813 (1996).

\bibitem{HDehnen1} H. Dehnen, H. Frommert and F. Ghaboussi, {\it Int. J. Theor.
Phys.} {\bf 31}, 109 (1992).

\bibitem{HDehnen2} H. Dehnen and H. Frommert, {\it Int. J. Theor.
Phys.} {\bf 32}, 1135 (1993).

\bibitem{HDehnen3} J. L. Cervantes-Cota and H. Dehnen, {\it Nucl.
Phys.} {\bf B442}, 391 (1995).

\bibitem{HDehnen4} J. L. Cervantes-Cota and H. Dehnen, {\it Phys.
Rev.} {\bf D51}, 395 (1995).

\bibitem{Bergmann} P. G. Bergmann, {\it Int. J. Theor. Phys.} {\bf 1}, 25 (1968).

\bibitem{Wagoner} R. V. Wagoner, {\it Phys. Rev.} {\bf D1}, 3209 (1970).

\bibitem{Nord} K. Nordtvedt, Jr., {\it Astrophys. J.} {\bf 161}, 1059 (1970).

\bibitem{Steinh} P. J. Steinhardt and F. S. Accetta, {\it Phys. Rev. Lett.} {\bf 64}, 2740 (1990).

\bibitem{Linde} A. D. Linde, Particle Physics and Inflationary
Cosmology, in {\it Contemporary Concepts in Physics}, Vol. 5
(Harwood, Chur, 1990).

\bibitem{Amen2} L. Amendola, S. Capozziello, M. Litterio and F. Occhionero, {\it
Phys. Rev.} {\bf D45}, 417 (1992).

\bibitem{Inagaki} T. Inagaki, S. Nojiri and S. D. Odintsov, {\it
JCAP}\; {\bf 0506}, 010 (2005).

\bibitem{Amen3} L. Amendola, {\it Phys. Lett.} {\bf B301}, 175 (1993).

\bibitem{Capo1} S. Capozziello and G. Lambiase, {\it Gen. Rel. Grav.} {\bf 31}, 1005 (1999).

\bibitem{Capo2} S. Capozziello, G. Lambiase and H.-J. Schmidt, {\it Ann. Phys. (Leipzig) } {\bf 9}, 39 (2000).

\bibitem{Prasa1} A. R. Prasanna, {\it Phys. Lett.} {\bf A37}, 331 (1971).

\bibitem{Prasa2} A. R. Prasanna, {\it Lett. Nuovo Cim.} {\bf 6}, 420 (1973).

\bibitem{Novello1} M. Novello and J. M. Salim, {\it Phys. Rev.} {\bf D20}, 377 (1979).

\bibitem{Drum} I. T. Drummond and S. J. Hathrell, {\it Phys. Rev.} {\bf D22}, 343 (1980).

\bibitem{Acci1} A. J. Accioly, A. N. Vaidya and M. M. Som, {\it Phys. Rev.} {\bf D28}, 1853
(1983).

\bibitem{Novello2} M. Novello and H. Heintzmann, {\it Gen. Rel. Grav.} {\bf 16}, 535 (1984).

\bibitem{Souza} J. G. Souza, M. L. Bedran and B. Lesche, {\it Rev. Bras.
Fis.} {\bf 14}, 488 (1984).

\bibitem{Go} H. F. M. Goenner, {\it Found. Phys.} {\bf 14}, 865 (1984).

\bibitem{Acci2} A. J. Accioly and N. L. P. Pereira da Silva, {\it Prog. Theor. Phys.} {\bf 76}, 1179 (1986).

\bibitem{Turner} M. S. Turner and L. M. Widrow, {\it Phys. Rev.} {\bf D37}, 2743 (1988).

\bibitem{Novello3} M. Novello, L. A. R. Oliveira and J. M. Salim, {\it Class. Quantum Grav.} {\bf 7}, 51 (1990).

\bibitem{Acci3} A. J. Accioly, A. D. Azeredo, C. M. L. de Arag\~ao and H. Mukai, {\it Class. Quantum Grav.}
{\bf 14}, 1163 (1997).

\bibitem{Mohanty} S. Mohanty and A. R. Prasanna, {\it Nucl. Phys.} {\bf B526}, 501 (1998).

\bibitem{Tess} P. Teyssandier, {\it Ann. Fond. Broglie} {\bf 29}, 173 (2004).

\bibitem{Lafrance} R. Lafrance and R. C. Myers, {\it Phys. Rev.} {\bf D51}, 2584 (1995).

\bibitem{Prasa3} A. R. Prasanna and S. Mohanty, {\it Class. Quantum Grav.}
 {\bf 20}, 3023 (2003).

\bibitem{Solanki} S. K. Solanki {\it et al\/}., {\it Phys. Rev.} {\bf D69}, 062001 (2004).

\bibitem{Kost1} V. A. Kosteleck\'y and M. Mewes, {\it Phys. Rev. Lett.} {\bf 87}, 251304 (2001).

\bibitem{Kost2} V. A. Kosteleck\'y and M. Mewes, {\it Phys. Rev.} {\bf D66}, 056005 (2002).

\bibitem{Balakin1} A. B. Balakin, {\it Class. Quantum Grav.} {\bf 14}, 2881 (1997).

\bibitem{Balakin3} A. B. Balakin and J. P. S. Lemos, {\it Class. Quantum Grav.}
{\bf 18}, 941 (2001).

\bibitem{Balakin4} A. B. Balakin, R. Kerner and J. P. S. Lemos, {\it Class. Quantum Grav.}
{\bf 18}, 2217 (2001).

\bibitem{Balakin5} A. B. Balakin and J. P. S. Lemos, {\it Class. Quantum Grav.} {\bf
19}, 4897 (2002).

\bibitem{Hehl1} G. F. Rubilar,  Yu. N. Obukhov and F. W. Hehl, {\it Class. Quantum Grav.} {\bf 20}, L185 (2003).

\bibitem{Hehl2} Ya. Itin and F. W. Hehl, {\it Phys. Rev.} {\bf D68}, 127701 (2003).

\bibitem{Hehl3} F. W. Hehl and Yu. N. Obukhov, in {\it Lecture Notes in
Physics}, Vol. 562 (Springer, Berlin, 2001), p.~479.

\bibitem{Horn} G. W. Horndeski, {\it Arch. Rat. Mech. Anal.} {\bf 75}, 229
(1981).

\bibitem{MH} F. M\"uller-Hoissen, {\it Class. Quantum Grav.} {\bf 5}, L35 (1988).

\bibitem{OdinDE} S. Nojiri, S. D. Odintsov, M. Sasaki, {\it Phys. Rev.} {\bf D71}, 123509 (2005).

\bibitem{BL05} A. B. Balakin and J. P. S. Lemos, {\it Class. Quantum Grav.} {\bf 22}, 1867 (2005).

\bibitem{BZ05} A. B. Balakin and W. Zimdahl, {\it Phys. Rev.} {\bf D71}, 124014 (2005).

\bibitem{1BZ06} A. B. Balakin and A. E. Zayats, {\it Gravit. Cosmol.} {\bf 12}, 302 (2006).

\bibitem{2BZ06} A. B. Balakin and A. E. Zayats, {\it Phys. Lett.} {\bf B644}, 294 (2007).

\bibitem{BSZ07} A. B. Balakin, S. V. Sushkov and A. E. Zayats, {\it Phys. Rev.} {\bf D 75}, 084042 (2007).

\bibitem{BDZ07} A. B. Balakin, H. Dehnen and A. E. Zayats, arxiv: 0710.5070.

\bibitem{Bij} J. J. van der Bij and E. Radu, {\it Nucl. Phys.} {\bf B585}, 637 (2000).

\bibitem{Rubakov} V. Rubakov, {\it Classical Theory of Gauge
Fields} (Princeton Univ. Press, Princeton, 2002).

\bibitem{Mosel} U. Mosel, {\it  Fields, Symmetries and Quarks}, (McGraw-Hill,
Hamburg, 1989).

\bibitem{Akhiezer} A. I. Akhiezer and S. V. Peletminsky, {\it
Fields and Fundamental Interactions} (Taylor and Francis, London,
2002).

\bibitem{Maugin} A. C. Eringen and G. A. Maugin, {\it Electrodynamics of
Continua} (Springer, New York, 1989).

\bibitem{Roskies} R. Roskies, {\it Phys. Rev.} {\bf D15}, 1722 (1977).

\bibitem{Mills} R. Mills, {\it Phys. Rev. Lett.} {\bf 43}, 549 (1979).

\bibitem{Goldin} G. A. Goldin and V. M. Shtelen, {\it J. Phys.} {\bf A37}, 10711 (2004).

\bibitem{MTW} C. W. Misner, K. S. Thorne and J. A. Wheeler, {\it Gravitation}
(Freeman, San Francisco, 1973).

\bibitem{Weinberg} S. Weinberg, {\it Gravitation and Cosmology} (Wiley, New York, 1972).

\end{thebibliography}
\end{document}